# Low temperature thermal history reconstruction using apatite fission-track length distribution and apatite U-Th/He age

Ruxin Ding[a, b, *]

[a] School of Earth Science and Engineering, Sun Yat-sen University, Guangzhou 510275, China

[b] Guangdong Provincial Key Laboratory of Mineral Resources & Geological Processes, Guangzhou 510275, China

*Corresponding author: dingrux@mail.sysu.edu.cn

**Abstract**: Low temperature thermochronology plays a key role in the study of tectonic evolution of the upper crust. The general application of thermal history modelling of apatite fission-track analysis requires both the parameters of the apparent age together with the confined track-length distribution of the spontaneous tracks. However, obtaining length data is relatively easy and does not require either irradiation or LA-ICP-MS commonly used for determining the uranium content of the grains for age dating. This leads to a shorter laboratory process. For this purpose, based on apatite U-Th/He method, this paper attempts to decouple apatite fission-track age from apatite fission-track length, and then combine the lengths with the respective apatite U-Th/He age to model the thermal history. Therefore, experiments were designed and conducted using a new program "*Low-T Thermo*". Results of this modelling are presented from the following experiments: apatite fission-track age combined with apatite U-Th/He age; apatite fission-track confined track-length distribution plus apatite U-Th/He age.

The modelling precision using this method is related to the relative errors of the apatite U-Th/He ages and the helium diffusion model. This combination of apatite fission-track length and apatite U-Th/He ages has not been implemented before but is presented here as an alternative way of determining thermal histories without the addition of apatite fission-track ages.

Keywords: Thermal history modelling; Apatite U-Th/He; Apatite fission track

**1 Introduction**

Low-temperature thermochronology including fission-track and (U–Th)/He analyses is a widely used tool for investigating tectonics and surface processes. Thermal history modelling is a key component providing a quantitative analysis of the temperature evolution. Generally, the common methods of thermal history modelling are based on a combination of apatite fission-track (AFT) age and confined track- length distributions (e.g. Ketcham et al., 2000; Ketcham, 2005; Gallagher, 1995, 2012; Gallagher et al., 2005). However, obtaining length data is relatively easy and does not require either irradiation or LA-ICP-MS commonly used for determining the uranium content of the grains for age dating. This leads to a shorter laboratory process. For this purpose, based on apatite U-Th/He (AHe) method, this paper attempts to decouple AFT age from AFT length, and then combine AFT lengths with the respective AHe age, in order to model the thermal history and compare these new combinations.

## 2 Thermal history modelling method

Although sophisticated software such as AFTSolve (Ketcham et al., 2000), HeFTy (Ketcham, 2005) and QTQt (Gallagher, 2012) have been developed successfully to model thermal histories from fission-track, and U-Th/He analyses, these software are not available for further/secondary development at present. For testing the above ideas and adding new modelling procedures, a new comprehensive Mathematica® modelling software ("*Low-T Thermo*") for improving thermal history analysis of fission-track and U-Th/He has been developed. The details of the modelling procedure and algorithms are presented as follows.

### 2.1 Forward modelling

The forward modelling procedure of AFT and U-Th/He in this paper includes four steps. Firstly, the thermal history is discretized into 100 evenly spaced time steps. Secondly, the reduced length distribution of the confined tracks is calculated for the duration of the thermal history. Thirdly, the modeled lengths are transferred into a fission-track modeled age. Finally helium diffusion is calculated (Flowers et al., 2009) based on the fission-track density and helium diffusion data is transferred into a U-Th/He modeled age.

**AFT age and AFT length**

The forward modelling procedure of apatite fission-track analysis is similar to that of Ketcham et al. (2000) and Ketcham (2005). The fanning curvilinear fit annealing model of Ketcham et al. (2007) is used for C-axis projected track-lengths, assuming that $r_{mr0}$ value is 0.83. The equation for C-axis projected track-lengths is from Ketcham et al. (2007). The initial C-axis projected track length is 16.62 μm (Ketcham et al., 2009). We assume a minimum detectable length of 7.31μm for C-axis projected track-length (Donelick et al., 1999). The assumption of a normal distribution of C-axis projected track-lengths is used at all stages of annealing. The relationship of C-axis projected mean versus standard deviation is $\sigma_{c,mod} = 0.00615 \times l_{c,mod}^2 - 0.177194 \times l_{c,mod} + 1.829975$ which is obtained by fitting the data from Carlson et al. (1999) based on the projection model of Ketcham (2007) (Appendix: Ketcham et al., 2007). The fission-track modeled age is calculated by the cumulated track density in each time step divided by 0.893 (Ketcham et al., 2000). The conversion model from fission-track length to density presented by Donelick et al. (1999) is used here. The forward modelled age and AFT length distribution is statistically consistent with HeFTy (Ketcham, 2005). The Kolmogorov-Smirnov (K-S) test probability, i.e. p-value (Marsaglia et al., 2003), is calculated based on the C-axis projected confined track-length distribution.

**AHe age**

In this paper, the following formula and models were used, including the spherical diffusion equation (Carslaw and Jaeger 1959), Arrhenius formula (Reiners and Brandon, 2006), U-Th/He age calculation formula (Farley, 2002), alpha ejection correction model (Farley et al., 1996) and the radiation damage accumulation and annealing model (RDAAM) for apatite (Parameter set 2 in table 1 in Flowers et al, 2009). The finite difference method (Ketcham, 2005) was used to calculate U-Th/He modeled age from the thermal history. The forward modeled AHe age is statistically concordant with the AHe modeled age calculated by HeFTy (Ketcham, 2005).

**2.2 Inverse modelling**

In this inverse process, we used the Monte Carlo method to randomly search thermal histories (e.g., 10,000) where time-temperature points are not regularly distributed and can be randomly perturbed.

Using the ages and fission-track length (any combination of AHe age, AFT age, and AFT length) to model the thermal history, the minimum of K-S test p-value and equivalent p-values were taken by assuming the grain ages have a normal distribution

$$p\text{-}value = 1 - \int_{O-|O-M|}^{O+|O-M|} \frac{e^{-\frac{(x-O)^2}{2\sigma^2}}}{\sqrt{2\pi}\sigma} dx$$

(similar to Ketcham, 2005). Where $O$ is the measured age, $M$ is the modeled age and $\sigma$ is the standard deviation of measured age. The 1σ age standard deviation is equivalent to K-S test p-value 0.32.

The mean value of thermal histories was selected within a threshold equivalent p-value as the result of the thermal modelling. The threshold equivalent p-value can be set as 0.5, 0.32, 0.05 etc. according to the data set. Because the time-step size is variable, each selected thermal history was subdivided into 100 evenly spaced time steps to calculate the mean temperature value at each time node.

**3 Synthetic examples**

Two synthetic examples are presented. In these examples, a simple cooling model and a reheating model were respectively analysed (Fig. 1) using two defined boxes as geological constraints. All pathways must pass through the boxes and be monotonic between the boxes. The age error (1σ) of both AFT and AHe is assumed to be 10% of the age and surface temperature is 10 ℃ for every method; the radius of a spherical crystal was set at 100 μm, and the concentration of both U and $^{232}$Th at 100 ppm for U-Th/He.

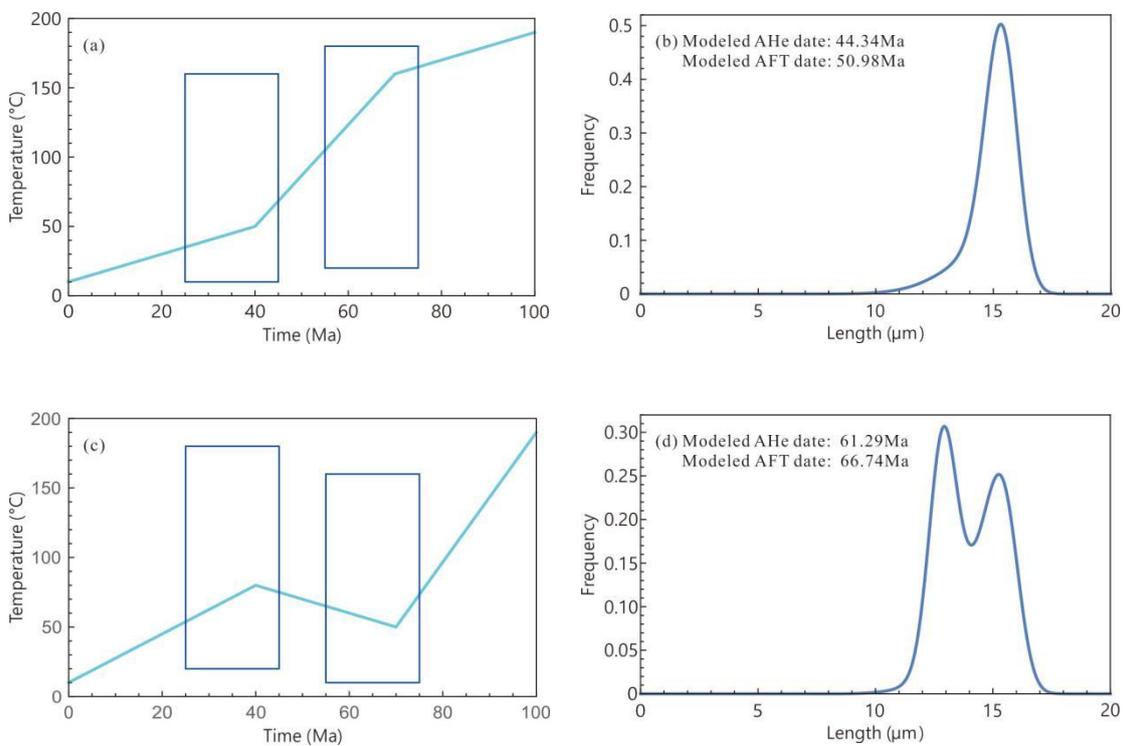

Fig.1 The input thermal histories and corresponding modeled ages and length distributions.

(a) and (b) represent the input thermal history and corresponding modeled ages and the C-axis projected length distribution in the monotonic model, respectively. (c) and (d) represent the input thermal history and corresponding modeled ages and the length distribution in the reheating model, respectively. The boxes represent the constraint ranges through which all thermal history paths must pass.

Four results are illustrated using the combination of apatite fission-track and U-Th/He data sets: a) AFT age and AFT length, b) AHe age and AFT age, c) AHe age and AFT length, d) AHe age and AFT age and AFT length (Fig. 2). All the acceptable thermal histories are within 1σ age error.

Above 110℃ there is a deviation where there is a separation between the modeled line and the input history, but under 110℃ the model results are good and close to the input history. The results show that decoupling the AFT age and AFT length separately combined with AHe age is possible.

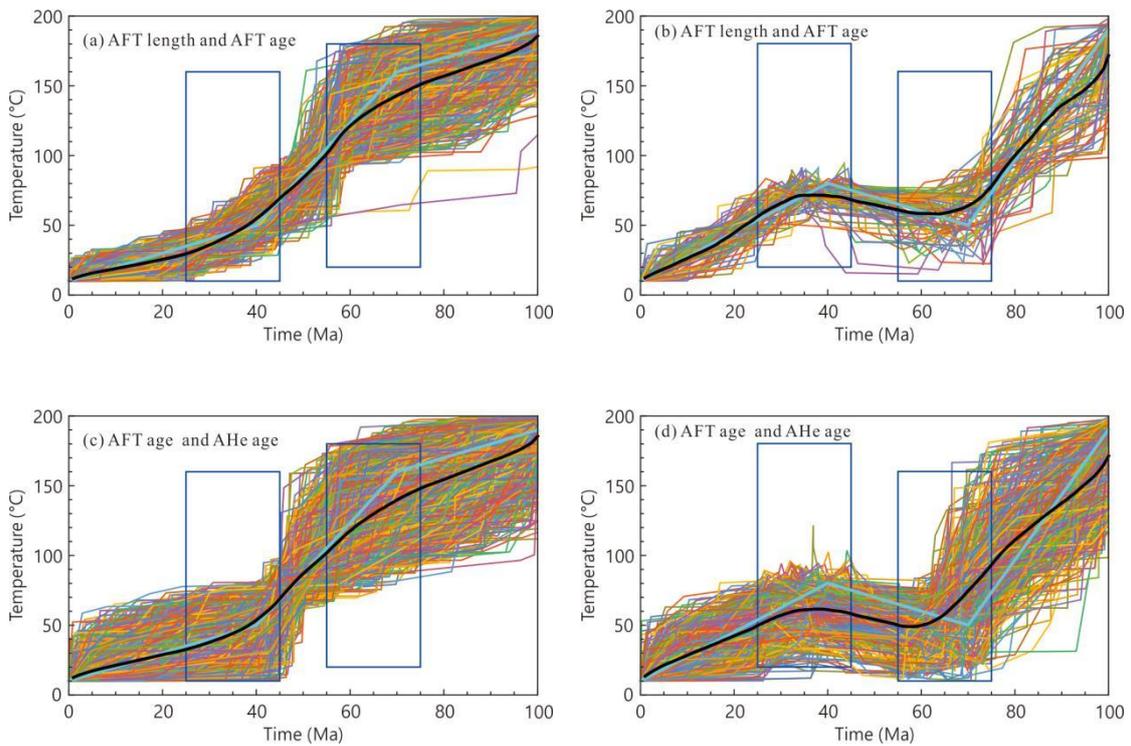

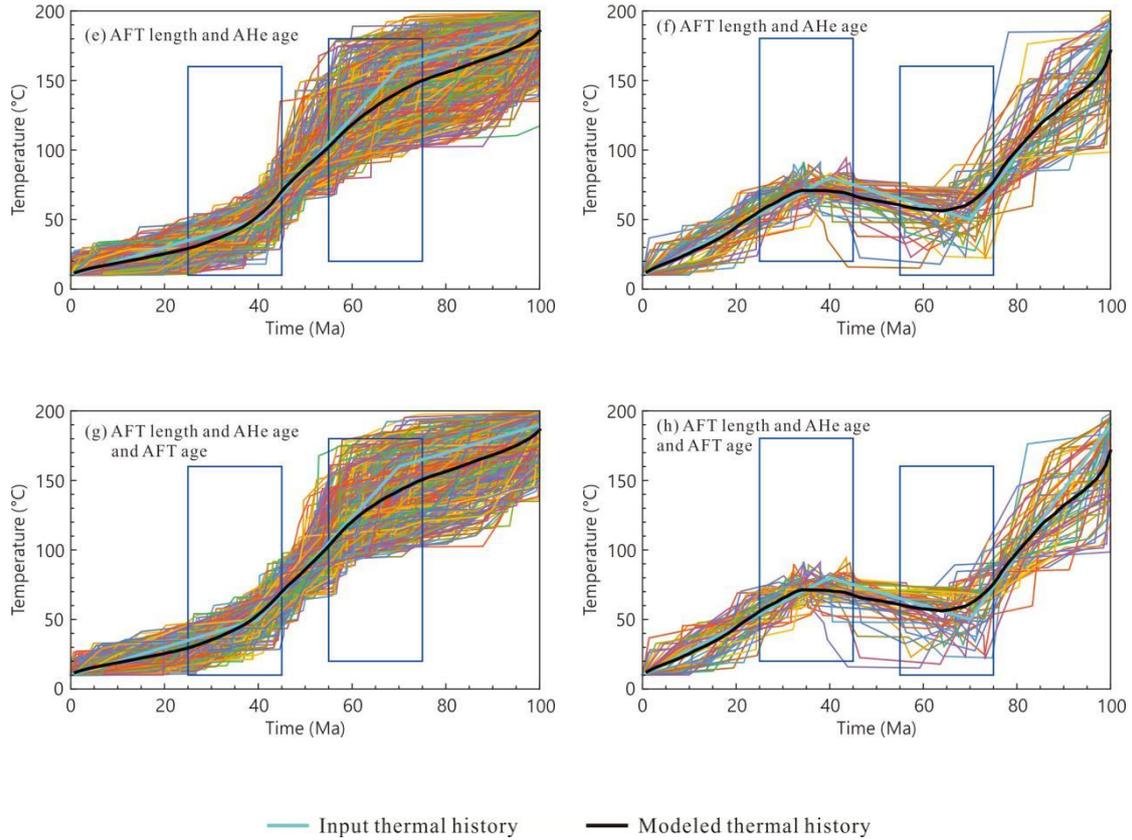

Fig.2 The thermal history modelling comparison of different combinations using AFT age, AFT length and AHe age for both the monotonic and the reheating models. All the thermal histories have ≥0.32 p-values (i.e., within 1σ age error).

## 4 Real examples

Examples are now presented to test the different combinations of AFT age, AFT length and AHe age. DB45 from Reiners et al. (2003) and Zhou et al. (2003), JR11-08 from Jiao et al. (2014) were chosen as test cases. DB45 is located in Tiantangzhai peak, Dabie Mountain, China, which experienced continuous exhumation since the Cretaceous. JR11-08 is from Ahimanawa Range, central North Island, New Zealand, where the basement rocks were exhumed to shallow depths of the crust in the

Early Cretaceous then followed by reheating before a second exhumation to shallow depths of the crust again.

Their age data are shown in Table 1 and C-axis projected length distributions in Figs. 3 and 4. The thermal histories of the combined AFT age, AHe age and AFT length were used as reference standards. The atmospheric lapse rate is assumed as 6 ℃/km. The present day surface temperature $T_s = T_{s0} - \beta \times h$, where $T_{s0}$ is the surface temperature at sea level, $\beta$ is the lapse rate and $h$ is the elevation.

The results (Figs. 3, 4) show that 1) both combinations of AFT age with AFT length and AHe age with AFT length are similar to that from the reference model; 2) the modeled temperature error range by combining AHe age with AFT length is smaller than that by combining AFT age with AFT length for DB45. On the contrary, the modeled temperature error range by combining AFT age with AFT length is a little less than that by combining AHe age with AFT length for JR11-08.

Table 1. Apatite (U-Th)/He data and apatite fission-track age data

| Sample Name | Elevation (m) | Grain No. | U (ppm) | Th (ppm) | $F_T$ | eU | AHe age (Ma) | 1σ (Ma) | Relative error | AFT age (Ma) | 1σ (Ma) | Relative error |
|---|---|---|---|---|---|---|---|---|---|---|---|---|
| DB45 | 340 | 1 | 10.5 | 3.5 | 0.73 | 11.32 | 22.5 | 0.7 | 3.11% | 47.4 | 6.6 | 13.92% |
| | | 1 | 28.3 | 64.9 | 0.76 | 43.55 | 25.8 | 1.9 | 7.36% | | | |
| JR11-08 | 431 | 2 | 6.5 | 15.0 | 0.74 | 10.03 | 15.8 | 1.2 | 7.59% | 83.9 | 5.2 | 6.20% |
| | | 3 | 24.6 | 34.4 | 0.72 | 32.68 | 31.7 | 2.4 | 7.57% | | | |

Note: DB45 is from Reiners et al. (2003) and Zhou et al. (2003), and JR11-08 is from Jiao et al. (2014).

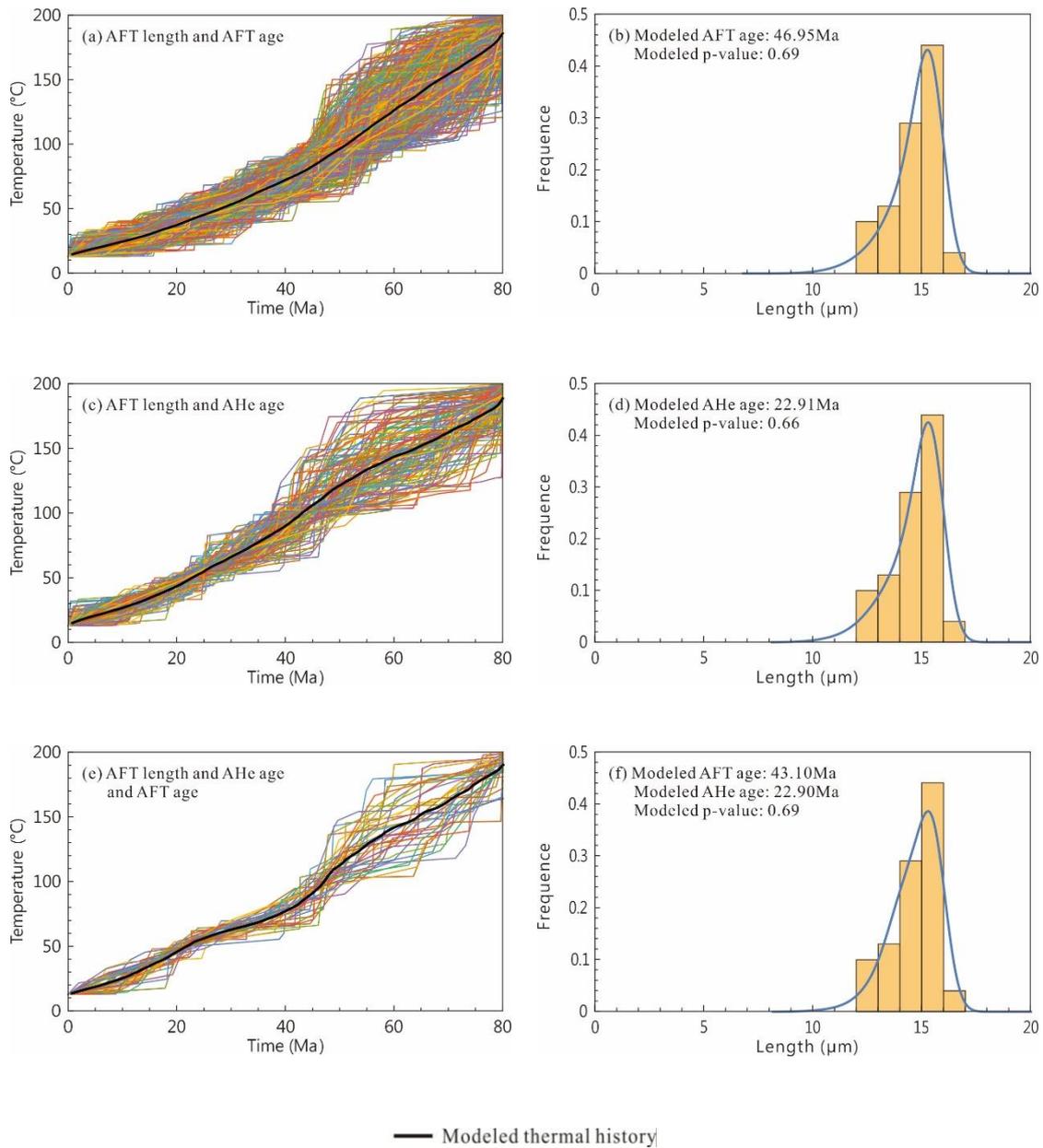

Fig.3 The thermal history modelling comparison of different combinations using AFT age, AFT length and AHe age for DB45.

All the thermal histories have ≥0.32 p-values (i.e., within 1σ age error). 10,000 thermal histories are used for the Monte Carlo randomly search. The surface temperature at sea level is 15℃. The atmospheric lapse rate is 6℃/km. The resulting models are very similar to those determined by Reiners et al. (2003) based on the AHe diffusion model (after Wolf et al., 1998) and AFTSolve.

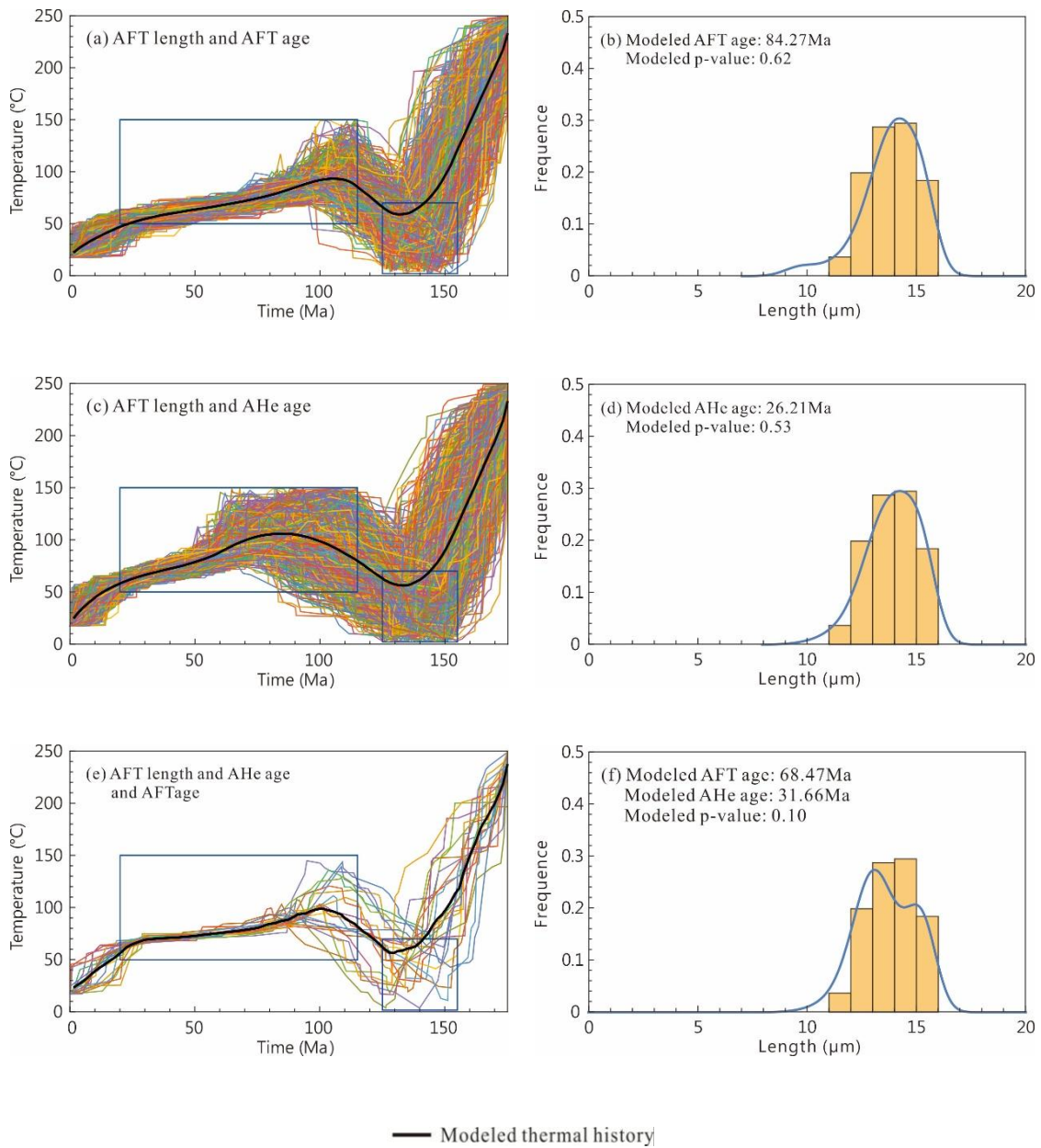

Fig.4 The thermal history modelling comparison of different combinations using AFT age, AFT length and AHe age of the first grain of JR11-08.

All the thermal histories in Fig.4a and Fig.4c have ≥0.05 p-values. All the thermal histories in Fig.4e have ≥0.001 p-values. 100,000 thermal histories are used for the Monte Carlo randomly search. The surface temperature at sea level is 20℃. The atmospheric lapse rate is 6℃/km. These models are very similar to those of Jiao et al. (2014) based on their use of QtQt.

## 5 Discussion

### 5.1 The constraints of AFT age and AHe age on AFT length

How might one estimate the constraints of AFT age and AHe age on AFT length in the thermal history modelling?

Firstly, using DB45 as the example (Fig. 5), one can see the difference of the thermal history modelling using AFT length, AFT age or AHe age respectively.

1) Although the thermal histories using only AFT length scatter on a larger time scale and anneal completely at ~110 ℃, the length modelling is sensitive to the whole temperature range between the AFT closure temperature and the surface temperature.

2) The thermal histories based separately on AHe age or AFT age intersect at or near the temperature of their partial retention zone (PRZ) or partial annealing zone (PAZ) and at the time of their respective apparent age, suggesting that the age modelling is more sensitive to the temperature of the respective PAZ or PRZ than outside these bounds. Therefore, the thermal history of both the AHe and AFT ages constrain the AFT length thermal history between surface temperature and AFT closure temperature (Fig. 5a, b, c). As expected, the AHe PRZ and AFT PAZ are the most sensitive temperature ranges for constraining AFT length thermal history.

3) For an AHe thermal history model using RDAAM, the temperature range between the AHe and AFT closure temperatures also has an influence on the AHe modeled age (Flowers et al., 2009). Therefore, the sensitive temperature range of the

AHe age thermal history modelling using RDAAM includes AHe PRZ and AFT PAZ.

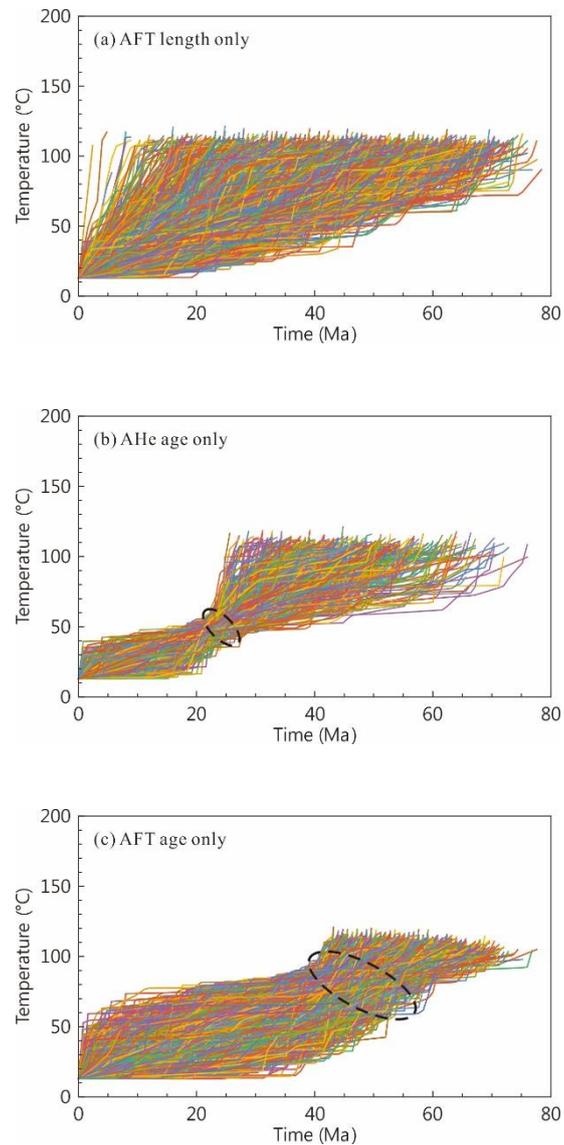

Fig.5 The thermal history modelling comparison using AFT length, AFT age or AHe age for DB45.

All the thermal histories have ≥0.32 p-values (i.e. within 1 σ age error). 10,000 thermal histories are used for the Monte Carlo randomly search. The thermal histories using only AFT length scatter and anneal completely at ~110 ℃. The thermal histories using only AFT age and AHe age cross at their respective partial annealing zones.

Secondly, the differences between the AHe age and AFT age relative errors (Table 1) also result in different constraints on AFT length thermal history modelling. These effects can be seen clearly from Figs 2 and 3. In Fig. 2 the relative errors of AHe age and AFT age are same. Therefore, the temperature error range of AFT age and AFT length thermal histories and that of AHe age and AFT length thermal histories are similar. However, in Fig. 3 the temperature error range of AHe age and AFT length thermal histories is smaller than that of AFT age and AFT length thermal histories because the relative error of the AHe age is smaller than that of the AFT age.

**5.2 The influence of multiple grain AHe ages**

AHe ages are known to frequently have poor reproducibility, so it is generally the practice to measure several grains from a single sample. How would such data be added to thermal history modelling? This is dependent on the helium diffusion model.

If the Durango fluorapatite model (Farley, 2000) is used, the age corresponding to a thermal history is constant and the age is not changed with the concentration of both U and $^{232}$Th. Therefore using the mean age of the analysed grains (e.g., the arithmetic mean age, central age (Vermeesch, 2010)) as the sample's AHe age, and combined with every grain's U, Th content and the alpha-ejection correction factor ($F_T$) each will yield a similar result.

However, if the RDAAM is selected as the helium diffusion model, it is difficult to give a sample's AHe age because the grain's age changes with eU (effective U concentration) and each grain has its own He closure temperature (Flowers et al., 2009). Therefore, the thermal history modelling results of different grains have different thermal sensitive ranges. Furthermore, the joint inversion of multiple grains improves the precision of the thermal history modelling. We use JR11-08 as a real example to test this viewpoint (Fig. 6).

As shown in Fig. 6, the joint inversion of 3 apatite grains has a more precise thermal history modelling result than any single grain AHe age. Further, the joint inversion of 3 apatite grains together with the AFT length distribution yields the most precise thermal history modelling result (Fig. 6e). It must be noted that an assumption was made that apatite U-Th zonation (Ault and Flowers, 2012), Bad Neighbours (Gautheron et al., 2012) and the crystal breakage (Brown et al., 2013) of all grains in the examples above have no significant influence on the thermal history modelling. The model results (Fig. 6e) are very similar to those determined by Jiao et al. (2014) based on their use of QtQt.

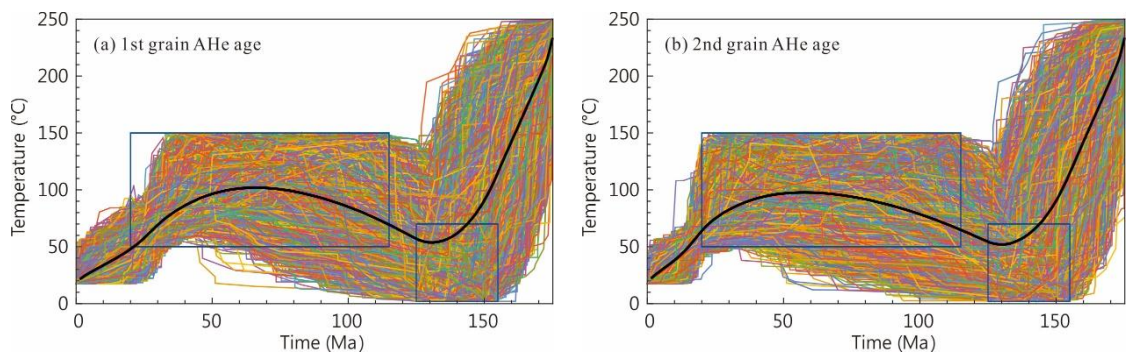

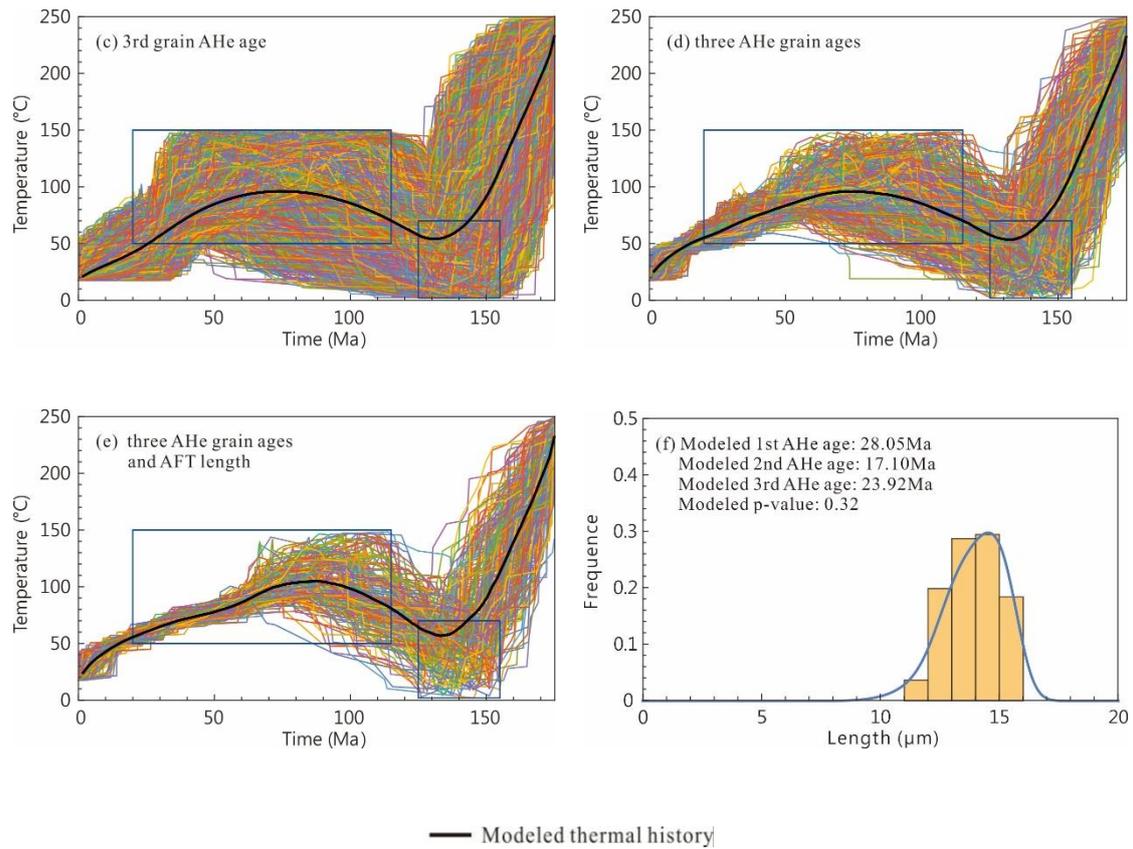

Fig. 6 The thermal history modelling comparison using different AHe grain-age data and confined track lengths from JR11-08. a-c JR11-08 thermal history modelling using grains 1, 2 and 3 respectively; d JR11-08 thermal history modelling based on combining three AHe grain ages and e, JR11-08 thermal history modelling based on combining three AHe grain ages and the AFT confined track lengths. All the thermal histories have ≥0.05 p-values. 100,000 thermal histories are used for the Monte Carlo random search. The minimum equivalent p-values is taken as the evaluating parameter. The surface temperature at sea level is 20 ℃. The atmospheric lapse rate is 6℃/km.

## 5 Conclusion

By testing the combination of AFT ages or AFT confined track lengths with AHe ages, it was determined that the decoupling of the apatite fission-track age and length is acceptable. The combination of AFT length and AHe ages yields results similar to other combinations with examples taken from the literature, and proves to be a very possible alternative method in modelling the data sets for thermal histories. Reconstruction of thermal histories offers more flexibility when AFT ages are not available. It may also act as a test for traditional modelling using the normally accepted procedure using a combination of AFT lengths and age. Certainly, when a higher quality modelling history is needed, both AHe age and AFT age are more recommendable.

## Acknowledgements

This study was supported by the National Natural Science Foundation of China (No. 41102131), the Fundamental Research Funds for the Central Universities of China (No. 12lgpy22), Guangdong Natural Science Foundation (No.2015A030313193) and China Scholarship Council. Diane Seward provided many important constructive suggestions for this paper writing, Shaowen Liu, Matt Sagar also provided reviewing this paper. The program is available through the author.